\documentclass[aps,prb, twocolumn, floatfix,10pt]{revtex4}
\usepackage{amssymb}
\usepackage{amsmath}
\usepackage{graphicx}
\usepackage{color}

\newcommand{\VL}{V_{\rm L}}
\newcommand{\VR}{V_{\rm R}}

\newcommand{\twelve}{$^{12}$C}
\newcommand{\thirteen}{$^{13}$C}
\newcommand{\boldtwelve}{$^{\mathbf{12}}$C}
\newcommand{\boldthirteen}{$^{\mathbf{13}}$C}
\newcommand{\bpar}{$B_{||}$}
\newcommand{\gs}{$\rm{g_s}$}

\begin{document}
\title{Electron-nuclear interaction in \thirteen~nanotube double quantum dots}
\author{H.~O.~H.~Churchill}
\affiliation{Department of Physics, Harvard University, Cambridge, Massachusetts 02138, USA}
\author{A.~J.~Bestwick}
\affiliation{Department of Physics, Harvard University, Cambridge, Massachusetts 02138, USA}
\author{J.~W.~Harlow}
\affiliation{Department of Physics, Harvard University, Cambridge, Massachusetts 02138, USA}
\author{F.~Kuemmeth}
\affiliation{Department of Physics, Harvard University, Cambridge, Massachusetts 02138, USA}
\author{D.~Marcos${\dagger}$}
\affiliation{Department of Physics, Harvard University, Cambridge, Massachusetts 02138, USA}
\author{C.~H.~Stwertka}
\affiliation{Department of Physics, Harvard University, Cambridge, Massachusetts 02138, USA}
\author{S.~K.~Watson${\dagger}$}
\affiliation{Department of Physics, Harvard University, Cambridge, Massachusetts 02138, USA}
\author{C.~M.~Marcus}
\affiliation{Department of Physics, Harvard University, Cambridge, Massachusetts 02138, USA}

\maketitle

{\bf For coherent electron spins, hyperfine coupling to nuclei in the host material can either be a dominant source of unwanted spin decoherence\cite{Khaetskii-PRL02,Petta-Sci05,Koppens-Nat06} or, if controlled effectively, a resource allowing storage and retrieval of quantum information\cite{Kane-Nat98,Taylor-PRL03,Dutt-Science07,Hanson-Science08}.
To investigate the effect of a controllable nuclear environment on the evolution of confined electron spins, we have fabricated and measured gate-defined double quantum dots with integrated charge sensors made from single-walled carbon nanotubes with a variable concentration of \boldthirteen~(nuclear spin $\mathbf{\emph I=1/2}$) among the majority zero-nuclear-spin \boldtwelve~atoms. 
Spin-sensitive transport in double-dot devices grown using methane with the natural abundance ($\mathbf{\sim 1\%}$) of  \boldthirteen~is compared with similar devices grown using an enhanced ($\mathbf{\sim 99\%}$) concentration of \boldthirteen.
We observe strong isotope effects in spin-blockaded transport and from the dependence on external magnetic field, estimate the hyperfine coupling in \boldthirteen~nanotubes to be on the order of $\mathbf{ 100~\mu}$eV, two orders of magnitude larger than anticipated theoretically\cite{Yazyev-NL08,Pennington-RMP96}.
 \boldthirteen-enhanced nanotubes are an interesting new system for spin-based quantum information processing and memory, with nuclei that are strongly coupled to gate-controlled electrons, differ from nuclei in the substrate, are naturally confined to one dimension, lack quadrupolar coupling, and have a readily controllable concentration from less than one to 10$^{5}$ per electron.
}
\footnotetext[0]{$\dagger$Present addresses:  Departamento de Teor\'{i}a de la Materia Condensada, Instituto de Ciencia de Materiales de Madrid, CSIC, Cantoblanco 28049, Madrid, Spain (D.M.); Department of Physics, Middlebury College, Middlebury, Vermont 05753, USA (S.K.W.).}

\begin{figure}[h!]
\center \label{figure1}
\includegraphics[width=3.4in]{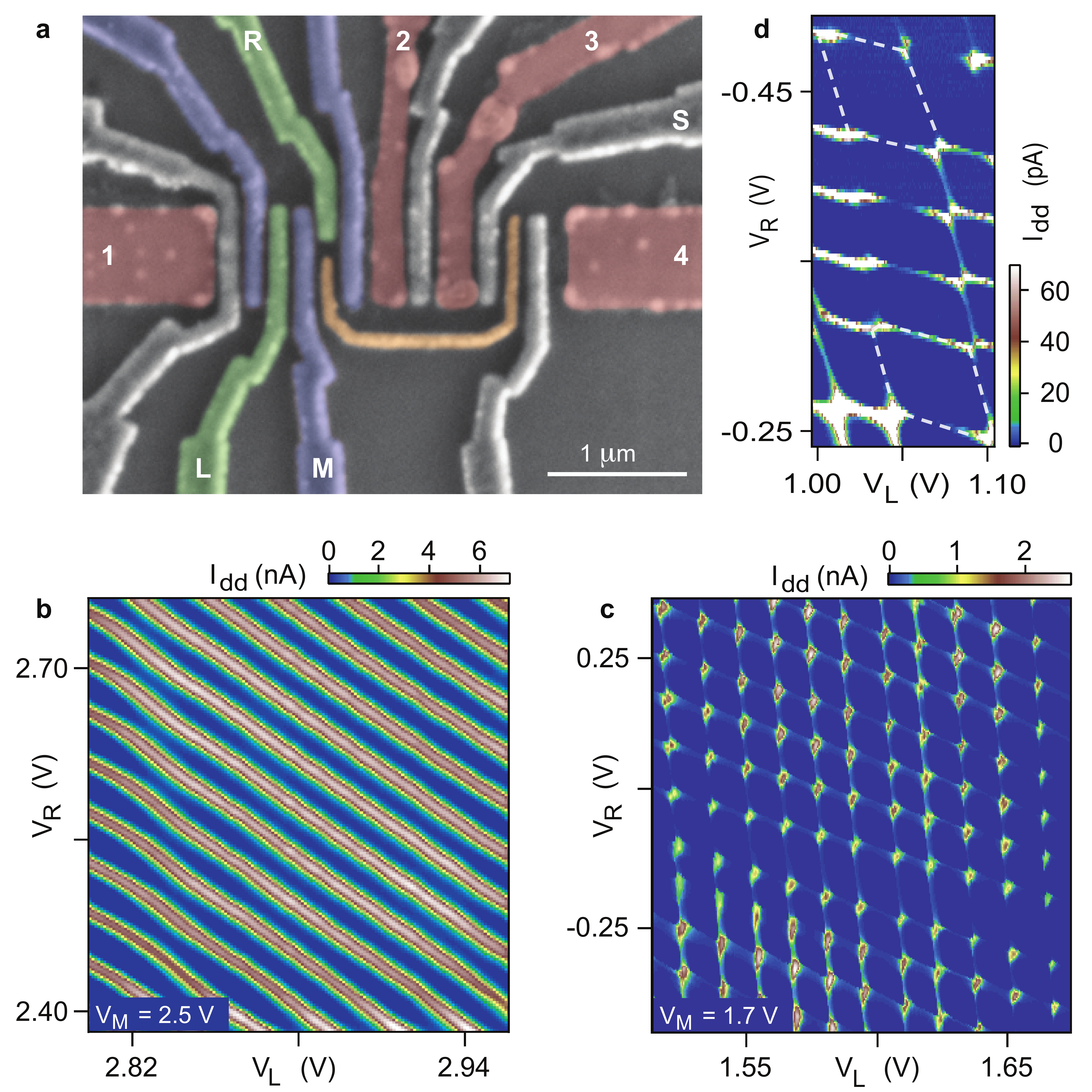}
\caption{\footnotesize{{\bf Nanotube double dot with integrated charge sensor.  a}, SEM micrograph (with false color) of a device similar to the measured \twelve~and \thirteen~devices.  The carbon nanotube (not visible) runs horizontally under the four Pd contacts (red).  Top-gates (blue) create voltage-tunable tunnel barriers allowing the formation of a single or double quantum dot between contacts 1 and 2.  Plunger gates L and R (green) control the occupancy of the double dot.  A separate single dot contacted by Pd contacts 3 and 4 is controlled with gate plunger gate S (gray) and is capacitively coupled to the double dot via a coupling wire (orange).  {\bf b}, Current through the double dot, $I_{\rm dd}$, (color scale) with the top-gates configured to form a large single dot.  {\bf c}, When carriers beneath the middle gate, M, are depleted, $I_{\rm dd}$ shows typical double-dot transport behavior, demarcating the honeycomb charge stability pattern. {\bf d}, Within certain gate voltage ranges, honeycomb cells with larger addition energy and fourfold periodicity (outlined with dashed lines) indicate the filling of spin and orbital states in shells. 
Source-drain bias is $-1.0$ mV for {\bf b}, {\bf c}, and {\bf d}.}}
\end{figure}

\par
Techniques to prepare, manipulate, and measure few-electron spin states in quantum dots has advanced considerably in recent years, with the leading progress in III-V semiconductor systems\cite{Ono-Sci02, Petta-Sci05,Koppens-Nat06, Hanson-RMP07}. 
All stable isotopes of III-V semiconductors, such as GaAs, have nonzero nuclear spin, and the hyperfine coupling of electron spins to host nuclei is a dominant source of spin decoherence in these materials\cite{Khaetskii-PRL02,Merkulov-PRB02,Petta-Sci05,Coish-PRB08}.
To eliminate this source of decoherence, group IV semiconductors---various forms of carbon, silicon, and silicon-germanium---which have predominantly zero nuclear spin, are being vigorously pursued as the basis of coherent spin electronic devices. 
Double quantum dots have recently been demonstrated in carbon nanotubes \cite{Biercuk-NL05,Sapmaz-NL06,Graeber-PRB06}, including recent investigation of spin effects \cite{Jorgensen-NPhys08,Buitelaar-PRB08}. 

\par 
The devices reported are based on single-walled carbon nanotubes grown by chemical vapor deposition using methane feedstock containing either 99\%~\thirteen~(denoted \thirteen~devices) or 99\% \twelve~(denoted \twelve~devices; see Methods)\cite{Liu-JACS01}.
The device design (Fig.~1a) uses two pairs of Pd contacts on the same nanotube;
depletion by top-gates (blue, green, and gray in Fig.~1a) forms a double dot between one pair of contacts and a single dot between the other. 
Devices are highly tunable, as demonstrated in Fig.~1, which shows that tuning the voltage on gate M (Fig.~1b) adjusts the tunnel rate between dots, allowing a cross-over from large single-dot behavior (Fig.~1b) to double-dot behavior  (Fig.~1c). Left and right tunnel barriers can be similarly tuned using the other gates shown in blue in Fig.~1a.

\par
A notable feature of nanotube quantum dots that is not shared by GaAs dots is that the energy required to add each subsequent electron, the addition energy, often shows shell-filling structure even in the many-electron regime\cite{Jorgensen-NPhys08}.
An example of a shell-filling pattern, with larger addition energy every fourth electron in the right dot, is seen in Fig.~1d. We find, however, that evident shell filling is not necessary to observe spin blockade at finite bias. Figures 2a and 2b show current through the double dot, $I_{\rm dd}$, as a function of gate voltages $\VR$~and $\VL$~for a weakly coupled, many-electron \thirteen~double dot at $+1$ and $-1$ mV source-drain bias, respectively, in a range of dot occupancy that does not show shell structure in the addition spectrum of either dot.  
With a magnetic field $B_{||}=200$~mT applied along the tube axis, current flow is observed throughout the finite-bias triangles at positive bias, but is suppressed at negative bias for detuning below 0.8 meV, which presumably indicates where an excited state of the right dot enters the transport window.

\par
Current rectification of this type is a hallmark of spin blockade\cite{Ono-Sci02} (Fig.~2e):  at positive bias, current flows freely as electrons of appropriate spin are drawn from the right lead to form the singlet ground state; at negative bias, current is blocked whenever a triplet state is formed between separated electrons, as the excess electron on the left can neither reenter the left lead nor occupy the lowest orbital state on the right without flipping its spin. Spin blockade was identified in all four devices measured, two each of \twelve~and \thirteen.
Spin blockade was occasionally found to follow a regular even-odd filling pattern, as seen in few-electron GaAs dots \cite{Johnson-PRB05}, though no pattern was seen adjacent to the area in Fig.~2.

\begin{figure}
\center \label{figure2}
\includegraphics[width=3.4in]{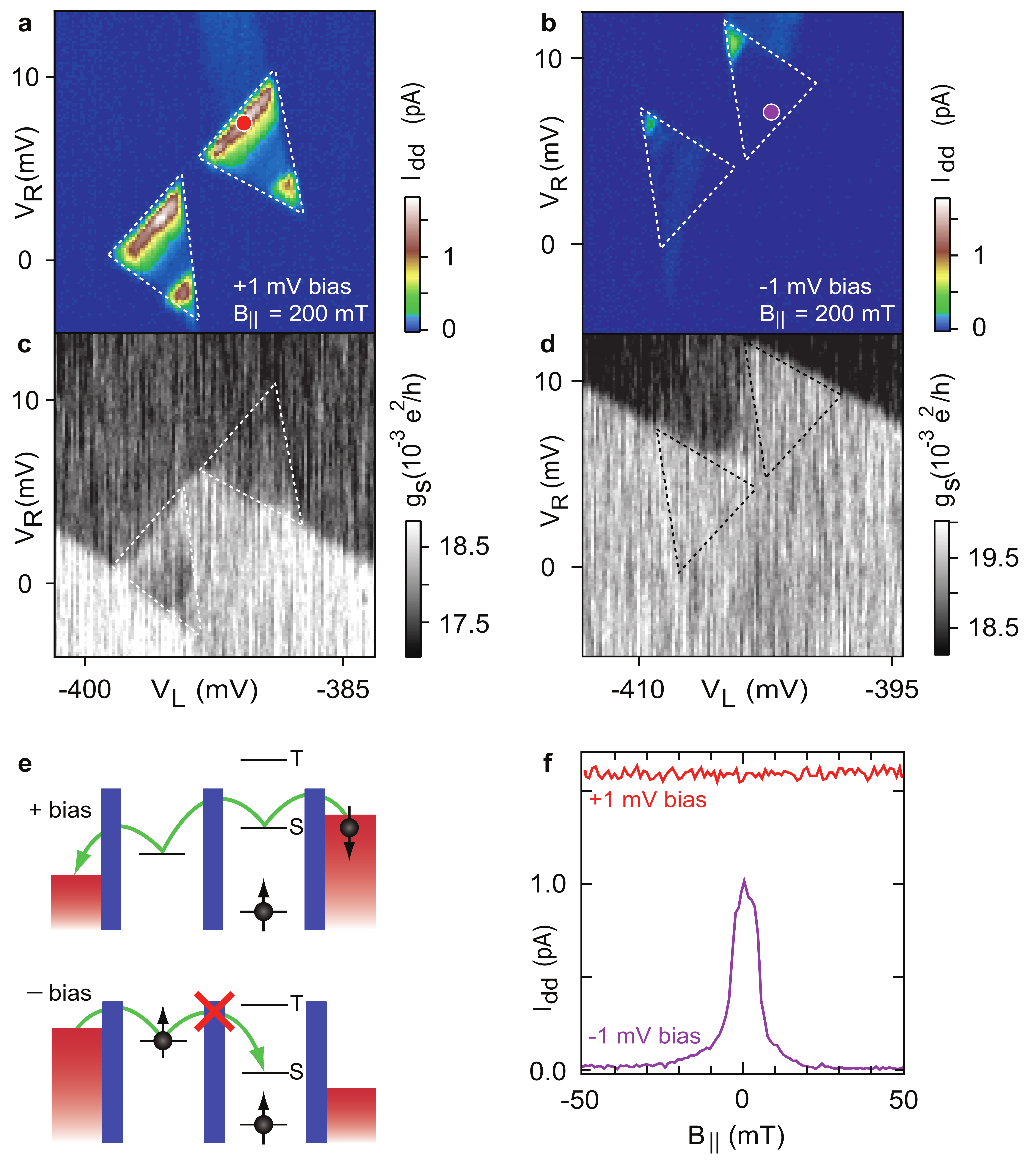}
\caption{\footnotesize{{\bf Spin blockade in a \boldthirteen~nanotube double dot.  a}, Current $I_{\rm dd}$ (color scale) at $+1.0$ mV source-drain bias, the non-spin-blockaded bias direction. Transport is dominated by resonant tunneling through the ground state at the base of the finite bias triangles and through an excited state at a detuning of 0.7 meV. {\bf b}, $I_{\rm dd}$ (color scale) at $-1.0$ mV source-drain bias, the spin-blockaded bias direction. $I_{\rm dd}$ is suppressed except near the tips of the transport triangles, where an excited state of the right dot becomes accessible. Suppressed transport for one bias direction is the signature of spin blockade. {\bf c}, Charge sensing signal, \gs, (conductance of the sensing dot between contacts 3 and 4 in Fig.~1a), acquired simultaneously with {\bf a} detects the the time-averaged occupation of the right dot.  {\bf d}, Charge sensing signal \gs~for $-1.0$ mV bias (blockade direction).  The transfer of charge from the left dot to the right is delayed until the excited state is reached at high detuning.  In {\bf a}--{\bf d} dashed lines indicate allowed regions for current flow in the absence of blockade.  {\bf e}, Schematic of spin-blockaded transport.  Any spin may occupy the left dot, but only a spin singlet is allowed in the right dot, suppressing negative bias current once an electron enters the left dot and forms a triplet state.
{\bf f}, Current $I_{\rm dd}$ at zero detuning as a function of magnetic field  for positive bias (non-blockade, red trace) and negative bias (blockade, purple trace).}}
\end{figure}

\par
Electrostatic sensing of the double-dot charge state is provided by a gate-defined quantum dot formed on a separately contacted portion of the same nanotube. The sensing dot is capacitively coupled to the double dot by a $\sim 1~\mu$m coupling wire\cite{Hu-NNano07} (orange gate in Fig.~1a) but electrically isolated by a depletion gate between the Pd contacts.
Charge sensor conductance ${\rm g_s}$ as a function of $\VR$~and $\VL$, acquired simultaneously with transport data in Fig.~2a,b, is shown in Fig.~2c,d. 
The location of the coupling wire makes ${\rm g_s}$ especially sensitive to occupancy of the right dot. 
Inside the positive-bias triangles (Fig.~2c), ${\rm g_s}$ is intermediate in value between their bordering regions, indicating that the excess electron is rapidly shuttling between the dots as current flows through the double dot.
In contrast, inside the negative-bias triangles (Fig.~2d), ${\rm g_s}$ shows no excess electron on the right dot as a result of spin blockade. These sensor values are consistent with models of finite-bias charge sensing in the spin-blockade regime \cite{Johnson-PRB05}.

\par
The magnetic field dependence of spin blockade provides important information about electron spin relaxation mechanisms\cite{Koppens-Sci05,Jouravlev-PRL06}.
A first look at field dependence (Fig.~2f) for a \thirteen~device shows that for negative bias (purple), spin-blockade leakage current is strongly peaked at ${B}_{||}=0$, while for positive bias, the unblockaded current does not depend on field. 
As discussed below, this field dependence can be understood in terms of hyperfine-mediated spin relaxation.

\begin{figure}
\center \label{figure3}
\includegraphics[width=3.4in]{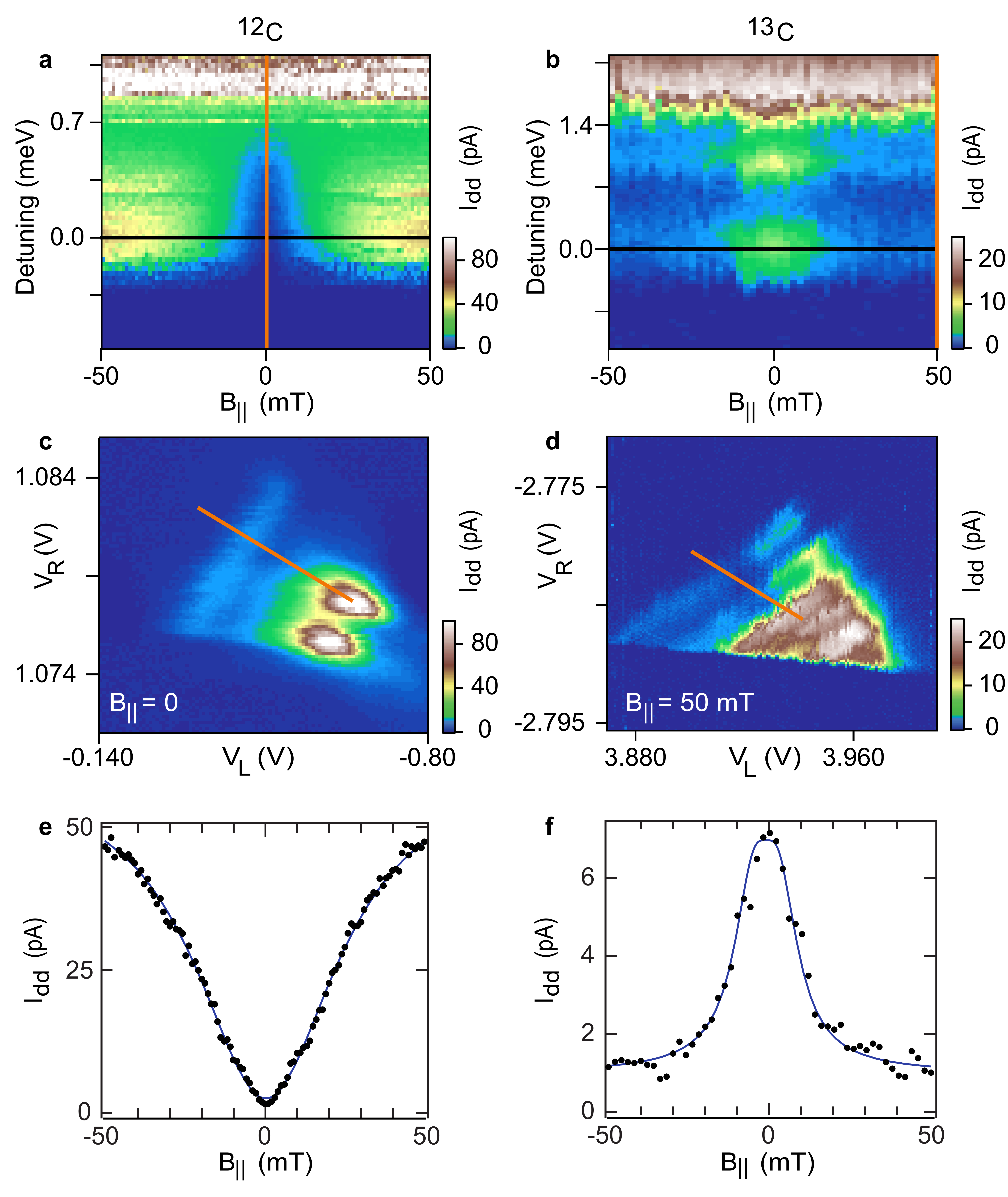}
\caption{\footnotesize{{\bf Contrasting magnetic field dependence of leakage current for \boldtwelve~and \boldthirteen~devices.}  Leakage current through spin blockade (color scale) as a function of detuning and magnetic field, \bpar, for ({\bf a}) \twelve~and ({\bf b}) \thirteen~devices.  The vertical axes in {\bf a} and {\bf b} are interdot detuning as indicated by the orange lines in {\bf c} and {\bf d}, respectively.  In {\bf a} \bpar~was swept and detuning stepped, while in {\bf b} detuning was swept and \bpar~stepped.  Bias is $-1.5$ mV in {\bf c} and $-4$ mV in {\bf d}.  {\bf e} and {\bf f} show cuts along \bpar~at the detunings indicated by the red lines in {\bf a} and {\bf b}, respectively.  The fit in {\bf e} is a Lorentzian with a width of 30 mT, and the fit in {\bf f} is to the theory of Jouravlev and Nazarov \cite{Jouravlev-PRL06}, providing a measure of $B_{\rm nuc}=6.1$ mT.}}
\end{figure}

\par
The striking difference in field dependence of spin-blockade leakage current between \twelve~and \thirteen~devices is illustrated in Fig.~3a,b. These data show that for negative (spin-blockaded) bias, leakage current is a minimum at $B_{||}=0$ for the \twelve~device  and a maximum at $B_{||}=0$ for the \thirteen~device. In fourteen instances of spin blockade measured in four devices (two \thirteen~and two \twelve), we find that leakage current minima can occur at $B_{||}=0$ in both \twelve~and \thirteen~devices, particularly for stronger interdot tunneling. For weaker interdot tunneling, however, only the \thirteen~devices show maxima of spin-blockade leakage at $B_{||}=0$. In all cases, the positive bias (non-spin-blockade) current shows no appreciable field dependence.

\par
Figure 3e shows spin-blockade leakage current as a function of \bpar~at fixed detuning (the detuning value is shown as a black line in Fig.~3a), along with a best-fit lorentzian, for the \twelve~device. The lorentzian form was not motivated by theory, but appears to fit rather well.
The width of the dip around $B_{||}=0$ increases with interdot tunneling (configuration Fig.~3e has $t\sim50~\mu$eV, based on charge-state transition width \cite{Hu-NNano07}). 
We note that a comparable zero-field dip in spin-blockade leakage current was recently reported in a double dot formed in an InAs nanowire \cite{Pfund-PRL07}, a material system with strong spin-orbit coupling. 
In the present system, the zero-field dip may also be attributable to spin-orbit coupling \cite{Kuemmeth-Nat08}, resulting in phonon-mediated relaxation that vanishes at $B_{||}=0$.

\par
Hyperfine coupling appears to the confined electrons as an effective local Zeeman field (the Overhauser field) that fluctuates in time independently in the two dots, driven by thermal excitation of nuclear spins. 
The difference in local Overhauser fields in the two dots will induce rapid mixing of all two-electron spin states whenever the applied field is less than the typical difference in fluctuating Overhauser fields. (At higher fields, only the $m=0$ triplet can rapidly mix with the singlet). How hyperfine-mediated spin mixing translates to a field dependence of spin-blockade leakage current was investigated experimentally in GaAs devices \cite{Koppens-Sci05}, with theory developed by Jouravlev and Nazarov. \cite{Jouravlev-PRL06} 

\par
Field dependence of spin-blockade leakage current for the \thirteen~device is shown Fig.~3f, along with a theoretical fit (Eq.~(11) of Ref.~\onlinecite{Jouravlev-PRL06}, with a constant background current added), from which we extract a root mean square amplitude of fluctuations of the local Overhauser fields, $B_{\rm nuc}=6.1$~mT.  
Assuming gaussian distributed Overhauser fields and uniform coupling, $B_{\rm nuc}$ is related to the hyperfine coupling constant $A$ by $g\mu_BB_{\rm nuc}=A/\sqrt{N},$ where $g$ is the electron g-factor and $N$ is the number of \thirteen~nuclei in each dot. \cite{Jouravlev-PRL06}
Taking $N\sim3$--$10 \times 10^4$ and $g=2$ (see Supplement), yields $A\sim1$--$2\times 10^{-4}$~eV, a value that is two orders of magnitude larger than predicted for carbon nanotubes \cite{Yazyev-NL08} or measured in fullerenes. \cite{Pennington-RMP96}

\begin{figure}
\center \label{figure4}
\includegraphics[width=3.4in]{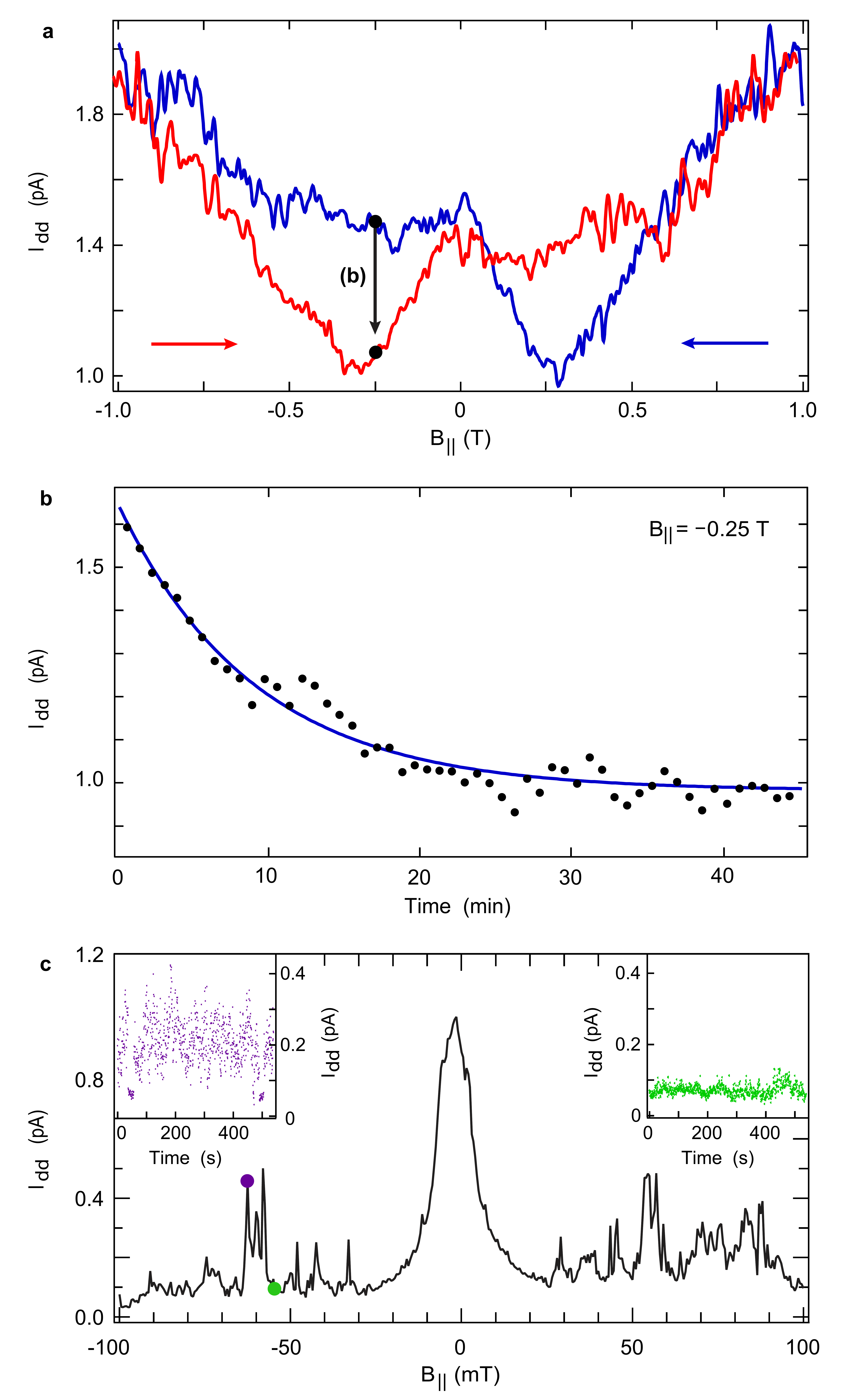}
\caption{\footnotesize{{\bf Hysteresis and fluctuations in leakage current.  a}, The spin-blockade leakage current for a \thirteen~device measured for decreasing (increasing) magnetic field (sweep rate 0.4 mT/s), shown in blue (red), after waiting at $+1$ T ($-1$ T) for 10~minutes. Hysteresis is seen on a field scale $>0.5$~T for both sweep directions.  {\bf b}, Decay of leakage current over time measured by stopping a downward sweep at $-0.25$ T.  The fit is to an exponential decay with a time constant of 9 min.  {\bf c}, Dependence of leakage current on \bpar~near zero detuning in a second \thirteen~device.  The leakage current fluctuates over time at some values of \bpar, while remaining steady at others (insets).}}
\end{figure}

\par
Signatures of dynamic nuclear polarization provide further evidence of a strong hyperfine interaction in \thirteen~double dots.
Hysteresis in the spin-blockade leakage current near zero detuning is observed when the magnetic field is swept over a tesla-scale range, as shown in Fig.~4a.  
The data in Fig.~4a,b are from the same \thirteen~device as in Fig.~3, but with the barriers tuned such that cotunneling processes provide a significant contribution to the leakage current.

\par 
We interpret the hysteresis in Fig.~4a as resulting from a net nuclear polarization induced by the electron spin flips required to circumvent spin blockade\cite{Baugh-PRL07}.
This nuclear polarization generates an Overhauser field felt by the electron spins that opposes \bpar~once \bpar~passes through zero.
The value of the coercive field, $B_c\sim0.6$ T, the external field at which the two curves rejoin, places a lower bound for the hyperfine
coefficient, $A\geq g\mu_BB_c\sim0.7 \times 10^{-4}$~eV (equality corresponding to full polarization), independent of the value inferred from the width of the leakage current peak around zero field (Fig.~3c).
If we instead use the value of $A$ inferred from the current peak width (Fig.~3c), the size of $B_c$ implies a  $\sim50\%$ polarization for the data in Fig.~4a.
Hysteresis is not observed for non-spin-blockaded transport in the \thirteen~devices and is not observed in the \twelve~devices, suggesting that this effect cannot be attributed to sources such as the Fe catalyst particles or interaction with nuclei in the substrate or gate oxide. 
\par 
Figure 4b shows that the induced nuclear polarization persists for $\sim10$~minutes, two orders of magnitude longer than similar processes in GaAs double dots \cite{Reilly-condmat08}. The long relaxation time indicates that nuclear spin diffusion is extremely slow, due both to the one-dimensional geometry of the nanotube and material mismatch between the nanotube and its surroundings. 
Field and occupancy dependence of relaxation were not measured.

\par
Large fluctuations in $I_{\rm dd}$ are seen at some values of magnetic field, but not at others (Fig.~4c), similar to behavior observed in GaAs devices \cite{Koppens-Sci05}. This presumably reflects an instability in nuclear polarization that can arise when polarization or depolarization rates themselves are polarization dependent \cite{Rudner-PRL07, Baugh-PRL07}.
\par 
An important conclusion of this work is that the hyperfine coupling constant, $A\sim 1$--$2 \times 10^{-4}$ eV, in the \thirteen~devices (for both electron and holes, see Methods) is much larger than anticipated \cite{Yazyev-NL08,Pennington-RMP96}.
It is possible that the substrate or gate oxide may enhance the degree of $s$-orbital content of conduction electrons,  thus strengthening the contact hyperfine coupling. We also note that a recent theoretical study of electron-nuclear spin interactions in \thirteen~nanotubes found that the one-dimensional character of charge carriers greatly enhances the effective electron-nuclear interaction \cite{Braunecker-condmat08}. Finally, we note that a large value of $A$ motivates the fabrication of isotopically enriched \twelve~nanotubes to reduce decoherence and the use of \thirteen~tubes as a potential basis of electrically addressable quantum memory. 

\section*{Methods}
Carbon nanotubes are grown by chemical vapor deposition using methane feedstock and 5 nm thick Fe catalyst islands on degenerately doped Si substrates with 1 $\mu$m thermal oxide.
\twelve~devices are grown with methane containing natural abundance (1.1\%) \thirteen; \thirteen~devices are grown with 99\% $^{13}$CH$_4$ (Sigma-Aldrich).  
Nanotubes are located after growth using a scanning electron microscope, and catalyst islands, source and drain electrodes (15 nm Pd), and top-gates (30 nm Al) are patterned using electron-beam lithography.
After contacting with Pd, samples are coated with a noncovalent functionalization layer combining NO$_2$ and trimethylaluminum, followed by atomic layer deposition (ALD) of a 30 nm Al$_2$O$_3$ top-gate insulator (Cambridge Nanotech Savannah ALD system)\cite{Farmer-NL06}.
Measurements were performed in a dilution refrigerator with a base temperature of 30 mK and electron temperature of $\sim 120$~mK, determined from the charge sensing transition width \cite{Hu-NNano07}.
Nanotubes presented in Figs.~1 and 2 have small bandgaps ($E_g\sim25$ meV); the \thirteen~nanotube in Fig.~3b,d,f and the other \twelve~nanotube (data not shown) are large-gap semiconducting nanotubes.
Charges occupying the dots and leads are electrons, except the data in Fig.~3b,d,f and Fig.~4a,b, where the charge carriers are holes. No significant differences are seen between devices with electron and hole carriers.

\vskip 0.25in
\noindent
{\bf Acknowledgements}
We thank Michael Biercuk, Karsten Flensberg, Leo Kouwenhoven, and Emmanuel Rashba for discussions, and David Reilly for experimental assistance. This work was supported in part by the National Science Foundation under grant no.~NIRT 0210736 and the NSF-NNIN Program, ARO/iARPA, the Department of Defense, Harvard's Center for Nanoscale Systems.  H.O.H.C.~acknowledges support from the NSF.
\ \\
\ \\
\noindent
{\bf Author Information}

Correspondence and requests for materials should be addressed to C.M.M.~(marcus@harvard.edu).


\small
\begin{thebibliography}{31}
\expandafter\ifx\csname natexlab\endcsname\relax\def\natexlab#1{#1}\fi
\expandafter\ifx\csname bibnamefont\endcsname\relax
\def\bibnamefont#1{#1}\fi
\expandafter\ifx\csname bibfnamefont\endcsname\relax
\def\bibfnamefont#1{#1}\fi
\expandafter\ifx\csname citenamefont\endcsname\relax
\def\citenamefont#1{#1}\fi
\expandafter\ifx\csname url\endcsname\relax
\def\url#1{\texttt{#1}}\fi
\expandafter\ifx\csname urlprefix\endcsname\relax\def\urlprefix{URL }\fi
\providecommand{\bibinfo}[2]{#2} \providecommand{\eprint}[2][]{\url{#2}}


\bibitem{Khaetskii-PRL02}
\bibinfo{author}{\bibfnamefont{A.~V.}~\bibnamefont{Khaetskii}}, \bibinfo{author}{\bibfnamefont{D.}~\bibnamefont{Loss}}, and \bibinfo{author}{\bibfnamefont{L.}~\bibnamefont{Glazman}}. \bibinfo{title}{Electron spin decoherence in quantum dots due to interaction with nuclei}.  \bibinfo{journal}{Phys.~Rev.~Lett.} \textbf{\bibinfo{volume}{88,}} \bibinfo{pages}{186802} (\bibinfo{year}{2002}).

\bibitem{Petta-Sci05}
\bibinfo{author}{\bibfnamefont{J.~R.}~\bibnamefont{Petta}}  {\it et al.}
\bibinfo{title}{Coherent manipulation of coupled electron spins in semiconductor quantum dots}. \bibinfo{journal}{Science} \textbf{\bibinfo{volume}{309,}} \bibinfo{pages}{2180} (\bibinfo{year}{2005}).

\bibitem{Koppens-Nat06}
\bibinfo{author}{\bibfnamefont{F.~H.~L.}~\bibnamefont{Koppens}}  {\it et al.}
\bibinfo{title}{Driven coherent oscillations of a single electron spin in a quantum dot}. \bibinfo{journal}{Nature} \textbf{\bibinfo{volume}{442,}} \bibinfo{pages}{766} (\bibinfo{year}{2006}).

\bibitem{Kane-Nat98}
\bibinfo{author}{\bibfnamefont{B.~E.}~\bibnamefont{Kane}}. \bibinfo{title}{A silicon-based nuclear spin quantum computer}.  \bibinfo{journal}{Nature} \textbf{\bibinfo{volume}{393,}} \bibinfo{pages}{133} (\bibinfo{year}{1998}).

\bibitem{Taylor-PRL03}
\bibinfo{author}{\bibfnamefont{J.~M.}~\bibnamefont{Taylor}}, \bibinfo{author}{\bibfnamefont{C.~M.}~\bibnamefont{Marcus}}, and \bibinfo{author}{\bibfnamefont{M.~D.}~\bibnamefont{Lukin}}.  \bibinfo{title}{Long-lived memory for mesoscopic quantum bits}.  \bibinfo{journal}{Phys.~Rev.~Lett.} \textbf{\bibinfo{volume}{90,}} \bibinfo{pages}{206803}  (\bibinfo{year}{2003}).
\bibitem{Dutt-Science07}
\bibinfo{author}{\bibfnamefont{M.~V.}~\bibnamefont{Gurudev Dutt}} {\it et al.}
\bibinfo{title}{Quantum register based on individual electronics and nuclear spin qubits in diamond}.  \bibinfo{journal}{Science} \textbf{\bibinfo{volume}{316,}} \bibinfo{pages}{1312}  (\bibinfo{year}{2007}).

\bibitem{Hanson-Science08}
\bibinfo{author}{\bibfnamefont{R.}~\bibnamefont{Hanson}}, \bibinfo{author}{\bibfnamefont{V.~V.}~\bibnamefont{Dobrovitski}}, \bibinfo{author}{\bibfnamefont{A.~E.}~\bibnamefont{Feiguin}}, \bibinfo{author}{\bibfnamefont{O.}~\bibnamefont{Gwyat}}, and \bibinfo{author}{\bibfnamefont{D.~D.}~\bibnamefont{Awschalom}}.  
\bibinfo{title}{Coherent dynamics of a single spin interacting with an adjustable spin bath}.  \bibinfo{journal}{Science} \textbf{\bibinfo{volume}{320,}} \bibinfo{pages}{352}  (\bibinfo{year}{2008}).

\bibitem{Yazyev-NL08}
\bibinfo{author}{\bibfnamefont{O.~V.}~\bibnamefont{Yazyev}}.  \bibinfo{title}{Hyperfine interactions in graphene and related carbon nanostructures}.  \bibinfo{journal}{Nano~Lett.}~\textbf{\bibinfo{volume}{8,}} \bibinfo{pages}{1011} (\bibinfo{year}{2008}).

\bibitem{Pennington-RMP96}
\bibinfo{author}{\bibfnamefont{C.~H.}~\bibnamefont{Pennington}} and \bibinfo{author}{\bibfnamefont{V.~A.}~\bibnamefont{Stenger}}.  \bibinfo{title}{Nuclear magnetic resonance of C$_{60}$ and fulleride superconductors}.  \bibinfo{journal}{Rev.~Mod.~Phys.}~\textbf{\bibinfo{volume}{68,}} \bibinfo{pages}{855} (\bibinfo{year}{1996}).

\bibitem{Ono-Sci02}
\bibinfo{author}{\bibfnamefont{K.}~\bibnamefont{Ono}}, \bibinfo{author}{\bibfnamefont{D.~G.}~\bibnamefont{Austing}}, \bibinfo{author}{\bibfnamefont{Y.}~\bibnamefont{Tokura}}, and \bibinfo{author}{\bibfnamefont{S.}~\bibnamefont{Tarucha}}. \bibinfo{title}{Current rectification by Pauli exclusion in a weakly coupled double quantum dot system}.
\bibinfo{journal}{Science} \textbf{\bibinfo{volume}{297,}} \bibinfo{pages}{1313} (\bibinfo{year}{2002}).

\bibitem{Hanson-RMP07}
\bibinfo{author}{\bibfnamefont{R.}~\bibnamefont{Hanson}} {\it et al.} 
\bibinfo{title}{Spins in few-electron quantum dots}.  \bibinfo{journal}{Rev.~Mod.~Phys.} \textbf{\bibinfo{volume}{79,}} \bibinfo{pages}{1217} (\bibinfo{year}{2007}).

\bibitem{Merkulov-PRB02}
\bibinfo{author}{\bibfnamefont{I.~A.}~\bibnamefont{Merkulov}}, \bibinfo{author}{\bibfnamefont{Al.~L.}~\bibnamefont{Efros}}, and \bibinfo{author}{\bibfnamefont{M.}~\bibnamefont{Rosen}}.  \bibinfo{title}{Electron spin relaxation by nuclei in semiconductor quantum dots}.  \bibinfo{journal}{Phys.~Rev.~B}~\textbf{\bibinfo{volume}{65,}} \bibinfo{pages}{205309} (\bibinfo{year}{2002}).

\bibitem{Coish-PRB08}
\bibinfo{author}{\bibfnamefont{W.~A.}~\bibnamefont{Coish}}, \bibinfo{author}{\bibfnamefont{J.}~\bibnamefont{Fischer}}, and \bibinfo{author}{\bibfnamefont{D.}~\bibnamefont{Loss}}.  \bibinfo{title}{Exponential decay in a spin bath}.  \bibinfo{journal}{Phys.~Rev.~B}~\textbf{\bibinfo{volume}{77,}} \bibinfo{pages}{125329} (\bibinfo{year}{2008}).

\bibitem{Biercuk-NL05}
\bibinfo{author}{\bibfnamefont{M.~J.}~\bibnamefont{Biercuk}}, \bibinfo{author}{\bibfnamefont{S.}~\bibnamefont{Garaj}}, \bibinfo{author}{\bibfnamefont{N.}~\bibnamefont{Mason}}, \bibinfo{author}{\bibfnamefont{J.~M.}~\bibnamefont{Chow}}, and \bibinfo{author}{\bibfnamefont{C.~M.}~\bibnamefont{Marcus}}. \bibinfo{title}{Gate-defined quantum dots on carbon nanotubes}. \bibinfo{journal}{Nano Lett.}~\textbf{\bibinfo{volume}{5,}} \bibinfo{pages}{1267} (\bibinfo{year}{2005}).

\bibitem{Sapmaz-NL06}
\bibinfo{author}{\bibfnamefont{S.}~\bibnamefont{Sapmaz}}, \bibinfo{author}{\bibfnamefont{C.}~\bibnamefont{Meyer}}, \bibinfo{author}{\bibfnamefont{P.}~\bibnamefont{Beliczynski}}, \bibinfo{author}{\bibfnamefont{P.}~\bibnamefont{Jarillo-Herrero}}, and \bibinfo{author}{\bibfnamefont{L.~P.}~\bibnamefont{Kouwenhoven}}. \bibinfo{title}{Excited state spectroscopy in carbon nanotube double quantum dots}.
\bibinfo{journal}{Nano Lett.} \textbf{\bibinfo{volume}{6,}} \bibinfo{pages}{1350} (\bibinfo{year}{2006}).

\bibitem{Graeber-PRB06}
\bibinfo{author}{\bibfnamefont{M.~R.}~\bibnamefont{Gr{\"{a}}ber}} {\it et al.} 
\bibinfo{title}{Molecular states in carbon nanotube double quantum dots}.
\bibinfo{journal}{Phys.~Rev.~B} \textbf{\bibinfo{volume}{74,}} \bibinfo{pages}{075427} (\bibinfo{year}{2006}).
\bibitem{Buitelaar-PRB08}
\bibinfo{author}{\bibfnamefont{M.~R.}~\bibnamefont{Buitelaar}} {\it et al.} 
\bibinfo{title}{Pauli spin blockade in carbon nanotube double quantum dots}.
\bibinfo{journal}{Phys.~Rev.~B} \textbf{\bibinfo{volume}{77,}} \bibinfo{pages}{245439} (\bibinfo{year}{2008}).

\bibitem{Jorgensen-NPhys08}
\bibinfo{author}{\bibfnamefont{H.~I.}~\bibnamefont{J{\o}rgensen}} {\it et al.} 
\bibinfo{title}{Singlet-triplet physics and shell filling in carbon nanotube double quantum dots}.
\bibinfo{journal}{Nature Phys.} \textbf{\bibinfo{volume}{4,}} \bibinfo{pages}{536} (\bibinfo{year}{2008}).

\bibitem{Liu-JACS01}
\bibinfo{author}{\bibfnamefont{L.}~\bibnamefont{Liu}} and \bibinfo{author}{\bibfnamefont{S.}~\bibnamefont{Fan}}.  \bibinfo{title}{Isotope labeling of carbon nanotubes and formation of $^{12}$C and $^{13}$C nanotube junctions}.  \bibinfo{journal}{J.~Am.~Chem.~Soc.} \textbf{\bibinfo{volume}{123,}} \bibinfo{pages}{11502}  (\bibinfo{year}{2001}).

\bibitem{Johnson-PRB05}
\bibinfo{author}{\bibfnamefont{A.~C.}~\bibnamefont{Johnson}}, \bibinfo{author}{\bibfnamefont{J.~R.}~\bibnamefont{Petta}}, \bibinfo{author}{\bibfnamefont{C.~M.}~\bibnamefont{Marcus}}, \bibinfo{author}{\bibfnamefont{M.~P.}~\bibnamefont{Hanson}}, and \bibinfo{author}{\bibfnamefont{A.~C.}~\bibnamefont{Gossard}}.
\bibinfo{title}{Singlet-triplet spin blockade and charge sensing in a few-electron double quantum dot}. \bibinfo{journal}{Phys.~Rev.~B} \textbf{\bibinfo{volume}{72,}} \bibinfo{pages}{165308} (\bibinfo{year}{2005}).

\bibitem{Hu-NNano07}
\bibinfo{author}{\bibfnamefont{Y.}~\bibnamefont{Hu}} {\it et al.} 
\bibinfo{title}{A Ge/Si heterostructure nanowire-based double quantum dot with integrated charge sensor}.
\bibinfo{journal}{Nature Nanotech.} \textbf{\bibinfo{volume}{2,}} \bibinfo{pages}{622} (\bibinfo{year}{2007}).

\bibitem{Jouravlev-PRL06}
\bibinfo{author}{\bibfnamefont{O.~N.}~\bibnamefont{Jouravlev}} and \bibinfo{author}{\bibfnamefont{Y.~V.}~\bibnamefont{Nazarov}}.  \bibinfo{title}{Electron transport in a double quantum dot governed by a nuclear magnetic field}.  \bibinfo{journal}{Phys.~Rev.~Lett.}~\textbf{\bibinfo{volume}{96,}} \bibinfo{pages}{176804} (\bibinfo{year}{2006}).

\bibitem{Koppens-Sci05}
\bibinfo{author}{\bibfnamefont{F.~H.~L.}~\bibnamefont{Koppens}} {\it et al.}
\bibinfo{title}{Control and detection of singlet-triplet mixing in a random nuclear field}.  
\bibinfo{journal}{Science}~\textbf{\bibinfo{volume}{309,}} \bibinfo{pages}{1346} (\bibinfo{year}{2005}).

\bibitem{Pfund-PRL07}
\bibinfo{author}{\bibfnamefont{A.}~\bibnamefont{Pfund}}, \bibinfo{author}{\bibfnamefont{I.}~\bibnamefont{Shorubalko}}, \bibinfo{author}{\bibfnamefont{K.}~\bibnamefont{Ensslin}}, and \bibinfo{author}{\bibfnamefont{R.}~\bibnamefont{Leturcq}}.
\bibinfo{title}{Suppression of spin relaxation in an InAs nanowire double quantum dot}. \bibinfo{journal}{Phys.~Rev.~Lett.} \textbf{\bibinfo{volume}{99,}} \bibinfo{pages}{036801} (\bibinfo{year}{2007}).

\bibitem{Kuemmeth-Nat08}
\bibinfo{author}{\bibfnamefont{F.}~\bibnamefont{Kuemmeth}}, \bibinfo{author}{\bibfnamefont{S.}~\bibnamefont{Ilani}}, \bibinfo{author}{\bibfnamefont{D.~C.}~\bibnamefont{Ralph}}, and \bibinfo{author}{\bibfnamefont{P.~L.}~\bibnamefont{McEuen}}. \bibinfo{title}{Coupling of spin and orbital motion of electrons in carbon nanotubes}.
\bibinfo{journal}{Nature} \textbf{\bibinfo{volume}{452,}} \bibinfo{pages}{448} (\bibinfo{year}{2008}).

\bibitem{Baugh-PRL07}
\bibinfo{author}{\bibfnamefont{J.}~\bibnamefont{Baugh}}, \bibinfo{author}{\bibfnamefont{Y.}~\bibnamefont{Kitamura}}, \bibinfo{author}{\bibfnamefont{K.}~\bibnamefont{Ono}}, and \bibinfo{author}{\bibfnamefont{S.}~\bibnamefont{Tarucha}}.
\bibinfo{title}{Large nuclear Overhauser fields detected in vertically coupled double quantum dots}. \bibinfo{journal}{Phys.~Rev.~Lett.} \textbf{\bibinfo{volume}{99,}} \bibinfo{pages}{096804} (\bibinfo{year}{2007}).

\bibitem{Reilly-condmat08}
\bibinfo{author}{\bibfnamefont{D.~J.}~\bibnamefont{Reilly}}  {\it et al.}
\bibinfo{title}{Exchange control of nuclear spin diffusion in a double quantum dot}. \bibinfo{journal}{cond-mat/0803.3082} (\bibinfo{year}{2008}).

\bibitem{Rudner-PRL07}
\bibinfo{author}{\bibfnamefont{M.~S.}~\bibnamefont{Rudner}}, and \bibinfo{author}{\bibfnamefont{L.~S.}~\bibnamefont{Levitov}}.
\bibinfo{title}{Self-polarization and dynamical cooling of nuclear spins in double quantum dots}. \bibinfo{journal}{Phys.~Rev.~Lett.} \textbf{\bibinfo{volume}{99,}} \bibinfo{pages}{036602} (\bibinfo{year}{2007}).


\bibitem{Braunecker-condmat08}
\bibinfo{author}{\bibfnamefont{B.}~\bibnamefont{Braunecker}}, \bibinfo{author}{\bibfnamefont{P.}~\bibnamefont{Simon}}, and \bibinfo{author}{\bibfnamefont{D.}~\bibnamefont{Loss}}.
\bibinfo{title}{Nuclear magnetism and Electronic Order in $^{13}$C nanotubes}. \bibinfo{journal}{cond-mat/0808.1685} (\bibinfo{year}{2008}).

\bibitem{Farmer-NL06}
\bibinfo{author}{\bibfnamefont{D.~B.}~\bibnamefont{Farmer}}, and \bibinfo{author}{\bibfnamefont{R.~G.}~\bibnamefont{Gordon}}.
\bibinfo{title}{Atomic layer deposition on suspended single-walled carbon nanotubes via gas-phase noncovalent functionalization}. \bibinfo{journal}{Nano Lett.} \textbf{\bibinfo{volume}{6,}} \bibinfo{pages}{699} (\bibinfo{year}{2006}).

\end{thebibliography}
\end{document}